# A General Euclidean Geometric Representation for the Classical Detection Theory

Muhammet Fatih Bayramoğlu
Department of Electrical and Electronics Engineering
Middle East Technical University
Ankara, Turkey
Email: fatih@eee.metu.edu.tr

Ali Özgür Yılmaz
Department of Electrical and Electronics Engineering
Middle East Technical University
Ankara, Turkey
Email: aoyilmaz@eee.metu.edu.tr

*Abstract*—We propose an Euclidean geometric representation for the classical detection theory. The proposed representation is so generic that can be employed to almost all communication problems. The hypotheses and observations are mapped into $\mathbb{R}^N$ in such a way that a posteriori probability of an hypothesis given an observation decreases exponentially with the square of the Euclidean distance between the vectors corresponding to the hypothesis and the observation.

## I. INTRODUCTION

One of the oldest and the most frequently employed tools in science is the Euclidean geometry. Visualizing a problem in an Euclidean geometry improves our understanding and usually contributes to the solution. Hence, Euclidean geometry plays a central role in many different disciplines such as astronomy, mathematics, and geography. The detection theory is not an exception.

The detection problem consists of a finite set of hypotheses, a set of observations, and a probabilistic transition mechanism between the hypotheses and the observations. Then the goal of the detection can be summarized as to guess the occurred hypothesis from the given observation [1].

In this paper we are interested in a particular case of the detection problem. That is detecting the transmitted symbol in the receiver of a communication system. In a communication system a symbol $X$ from a finite alphabet $\mathcal{A}$ is transmitted through an imperfect channel. The channel stochastically maps this transmitted symbol to an output $Y$ which is an element of an observation space $\mathcal{R}$. $\mathcal{R}$ may consist of a finite number of symbols, real numbers, or functions depending on the problem. The goal of the receiver is to predict $X$ given $Y$.

What we mean by an Euclidean geometric representation of the detection problem is mapping the alphabet $\mathcal{A}$ and the observation space $\mathcal{R}$ to $\mathbb{R}^N$ which is a well known *finite dimensional* Euclidean space. This mapping of the symbols and observations should posses some properties in order to be useful. First of all, this mapping should be in such a way that the Euclidean distance between two vectors representing a symbol and an observation should reflect the probabilistic transition mechanism of the detection problem. In a communication system we usually transmit a stream of symbols through a channel rather than transmitting just a single symbol. Therefore, this representation should easily be extended to the multiple usages of the channel. Finally the translation of the vectors should have a useful meaning.

The Euclidean geometry is employed in the communication theory for a particular problem which is communication through additive white Gaussian noise (AWGN) channels. The signal space representation posses many nice properties that it became the standard representation for explaining this problem [1], [2], [3]. However, the application of the signal space representation is restricted to the detection under AWGN.

Our goal in this paper is to develop a general Euclidean geometric representation for the detection theory which fits to almost all of the communication problems, such as non-Gaussian noise channels or even discrete output channels as well as AWGN channels. Since we would like our representation to be a generic one, we have to sacrifice from some of the properties that signal space representation has. However, the developed representation still shares certain important properties of the signal space representation.

In this work our aim is to provide a geometric viewpoint for the detection problem. Hence, we are not concerned with the computation of the a posteriori probabilities or likelihoods and we assume that these values are known.

While mapping $\mathcal{R}$ to $\mathbb{R}^N$ we make use of the log-likelihood ratio which is a very well known concept in communication theory. In other words, we show that log-likelihood ratio is not only a parameter for optimum decision but also provides a full geometric representation for the observations.

Developing a geometric representation for detection theory is first attempted by Dabak in [4]. However, the geometry he proposed is not Euclidean, not even Riemannian. On the other hand, the geometry that we propose is Euclidean.

This paper is organized as follows. In the next section the desired properties from a geometric representation are explained. The proposed geometric representation is introduced in the third section. The fourth section provides some examples and then the paper is concluded.

Throughout the paper we adopt the following notation. We denote deterministic variables by lowercase letters whereas random variables by uppercase letters except letter $C$ which is employed in denoting the normalization constants. Boldface letters denote vectors. Finally sets are denoted by calligraphic letters except $\mathbb{R}^N$.

## II. THE PROPERTIES EXPECTED FROM AN EUCLIDEAN GEOMETRIC REPRESENTATION OF THE COMMUNICATION PROBLEM

We need two operators and a real function to formally describe an Euclidean geometric representation of a communication channel. The first operator is required to map the symbols in the alphabet $\mathcal{A}$ to the vectors in $\mathbb{R}^N$. This operator is denoted by $\mathcal{M}_\mathcal{A}\{.\}$. The second operator is used to map the observations in $\mathcal{R}$ to $\mathbb{R}^N$ and denoted by $\mathcal{M}_\mathcal{R}\{.\}$. Finally we need a function $f(.)$ from $\mathbb{R}^+ \cup \{0\}$ to $\mathbb{R}^+$ to describe the a posteriori probability of the transmitted symbol $X$ given the observation $Y$ in terms of the distance between $\mathcal{M}_\mathcal{A}\{X\}$ and $\mathcal{M}_\mathcal{R}\{Y\}$ as follows.

$$\Pr\{X = x|Y = y\} = \frac{1}{C} f\left(\|\mathcal{M}_\mathcal{R}\{y\} - \mathcal{M}_\mathcal{A}\{x\}\|\right) \quad (1)$$

where $\|.\|$ denotes the usual norm in $\mathbb{R}^N$ and the normalization constant $C$ is given by

$$C = \sum_{\forall x \in \mathcal{A}} f(\|\mathcal{M}_\mathcal{R}\{Y\} - \mathcal{M}_\mathcal{A}\{x\}\|)$$

The most important requirement to be imposed on the geometric representation is related to the function $f(.)$. The function $f(.)$ should be a monotonically decreasing function. In other words $\Pr\{X|Y\}$ should decrease as the distance between $\mathcal{M}_\mathcal{A}\{X\}$ and $\mathcal{M}_\mathcal{R}\{Y\}$ increases.

As it is mentioned in the introduction part, we usually transmit a sequence of symbols through a communication channel and the geometric representation should be in consistent with the multiple usage of the channel. Suppose that a vector $\mathbf{X} = [X_1, X_2, \ldots, X_M]$ is to be transmitted through the channel by using it $M$ times. We assume that channel is memoryless. Let the observations be denoted by $\mathbf{Y} = [Y_1, Y_2, \ldots, Y_M]$. The combined alphabet in this problem is $\mathcal{A}^M$ and the observation space is $\mathcal{R}^M$. We need two operators from these alphabet and observation space to $\mathbb{R}^{NM}$. In order to keep the consistency with the single usage of the channel, these operators can be defined simply by concatenating the $\mathcal{M}_\mathcal{A}\{X_i\}$ and $\mathcal{M}_\mathcal{R}\{Y_i\}$ for $i = 1, 2, \ldots, M$ as follows.

$$\mathcal{M}_{\mathcal{A}^M}\{\mathbf{X}\} \triangleq \begin{bmatrix} \mathcal{M}_\mathcal{A}\{X_1\} \\ \mathcal{M}_\mathcal{A}\{X_2\} \\ \vdots \\ \mathcal{M}_\mathcal{A}\{X_M\} \end{bmatrix} \quad (2)$$

$$\mathcal{M}_{\mathcal{R}^M}\{\mathbf{Y}\} \triangleq \begin{bmatrix} \mathcal{M}_\mathcal{R}\{Y_1\} \\ \mathcal{M}_\mathcal{R}\{Y_2\} \\ \vdots \\ \mathcal{M}_\mathcal{R}\{Y_M\} \end{bmatrix} \quad (3)$$

For the sake of consistency we desire that the a posteriori probability of $\mathbf{X}$ given $\mathbf{Y}$ is related to $\|\mathcal{M}_{\mathcal{R}^M}\{\mathbf{Y}\} - \mathcal{M}_{\mathcal{A}^M}\{\mathbf{X}\}\|$ with the same function $f(.)$ as in the single usage case. In other words we would like to have

$$\Pr\{\mathbf{X}|\mathbf{Y}\} = \frac{1}{C_1} f(\|\mathcal{M}_{\mathcal{R}^M}\{\mathbf{Y}\} - \mathcal{M}_{\mathcal{A}^M}\{\mathbf{X}\}\|) \quad (4)$$

where $C_1$ is a normalization coefficient. Since, the norm appearing in the equation above is an *Euclidean* norm

$$\Pr\{\mathbf{X}|\mathbf{Y}\} = \frac{1}{C_1} f\left(\sqrt{\sum_{i=1}^{M} \|\mathcal{M}_\mathcal{R}\{Y_i\} - \mathcal{M}_\mathcal{A}\{X_i\}\|^2}\right) \quad (5)$$

Since, we assumed the that channel is memoryless

$$\begin{aligned} \Pr\{\mathbf{X}|\mathbf{Y}\} &= \prod_{i=1}^{M} \Pr\{X_i|Y_i\} \\ &= \frac{1}{C_2} \prod_{i=1}^{M} f(d_i) \quad (6) \end{aligned}$$

where $d_i = \|\mathcal{M}_\mathcal{R}\{Y_i\} - \mathcal{M}_\mathcal{A}\{X_i\}\|$. Combining this result with (5) imposes a further condition on $f(.)$. That is $f(.)$ should satisfy the following equation

$$f\left(\sqrt{\sum_{i=1}^{M} d_i^2}\right) = \alpha \prod_{i=1}^{M} f(d_i) \quad (7)$$

for all positive $d_i$ and an $\alpha$ in $\mathbb{R}$. There is only a single form of function which satisfies this requirement. In order to satisfy this requirement $f(.)$ is necessarily in the form of

$$f(x) = \beta \exp(\gamma x^2) \quad (8)$$

where $\beta$ is a positive real number. Recall that the function $f(.)$ should be a monotonically decreasing function. Hence, $\gamma$ should be a negative real number.

In order to make the Euclidean geometric representation more useful we should assign a useful meaning to the translation (vector addition in $\mathbb{R}^N$) operation. In some specific communication channels, such as AWGN channel, the observation space $\mathcal{R}$ has well defined addition and scaling operations. If this was true for all communication channels then we could assign a meaning to the translation operation by imposing a linearity constraint on the operator $\mathcal{M}_\mathcal{R}\{.\}$. However, there are many channels with an observation space which do not posses a well defined addition operation. Therefore, we have to adopt a more general approach for assigning a meaning to the translation operation.

We assign a meaning to translation operation related to the multiple usage of the channel again. However, in this case we assume that $M$ *replicas* of an *equally likely distributed* symbol $X$ is transmitted through the channel instead of an arbitrary vector $\mathbf{X}$. Let the observations be denoted by $\mathbf{Y} = [Y_1, Y_2, \ldots, Y_M]$. A straight forward approach for developing an Euclidean geometric representation for this problem would be using the operators defined in (2) and (3) with $\mathbf{X} = [X, X, \ldots, X]$. However, a better representation can be developed by making use of the fact that $M$ replicas of the same symbol is transmitted in this case. We propose the following operator from $\mathcal{R}^M$ to $\mathbb{R}^N$ (rather than $\mathbb{R}^{NM}$) to map $\mathbf{Y}$ resulting from transmitting $M$ replicas of an equally likely distributed symbol through the channel.

$$\mathcal{M}_{\mathcal{R}^M}\{\mathbf{Y}\} \triangleq \sum_{i=1}^{M} \mathcal{M}_{\mathcal{R}}\{Y_i\} \quad (9)$$

Then the translation operation may become useful if the a posteriori probability of $X$ given $\mathbf{Y}$ can be expressed as

$$\begin{aligned}\Pr\{X|\mathbf{Y}\} &= f(\|\mathcal{M}_{\mathcal{R}^M}\{\mathbf{Y}\} - \mathcal{M}_{\mathcal{A}}\{X\}\|) \\ &= f\left(\left\|\sum_{i=1}^{M} \mathcal{M}_{\mathcal{R}}\{Y_i\} - \mathcal{M}_{\mathcal{A}}\{X\}\right\|\right)\end{aligned} \quad (10)$$

## III. THE TRANSFORMATIONS FROM $\mathcal{A}$ AND $\mathcal{R}$ TO $\mathbb{R}^N$

The necessary properties that an Euclidean geometric representation should posses are laid down in the previous section. In this section, we propose the operators $\mathcal{M}_{\mathcal{A}}\{.\}$, $\mathcal{M}_{\mathcal{R}}\{.\}$ and then prove that these operators satisfy conditions imposed in the previous section.

First of all the codomain of $\mathcal{M}_{\mathcal{A}}\{.\}$ and $\mathcal{M}_{\mathcal{R}}\{.\}$ should be more clearly specified. The input symbols and channel observations are mapped to $\mathbb{R}^N$ by $\mathcal{M}_{\mathcal{A}}\{.\}$ and $\mathcal{M}_{\mathcal{R}}\{.\}$ where $N$ is the *number of elements in the alphabet* $\mathcal{A}$.

We need to enumerate the input symbols before defining $\mathcal{M}_{\mathcal{A}}\{.\}$. Let the elements of $\mathcal{A}$ be $\{x_1, x_2, \ldots, x_N\}$. Then $\mathcal{M}_{\mathcal{A}}\{.\}$ is defined from $\mathcal{A}$ to $\mathbb{R}^N$ as

$$\mathcal{M}_{\mathcal{A}}\{x_i\} \triangleq \frac{1}{2N}\mathbf{e}_i - \frac{1}{2N^2}\sum_{j=1}^{N}\mathbf{e}_j \quad (11)$$

where $\mathbf{e}_i$ is the $i^{th}$ canonical basis of $\mathbb{R}^N$. The norm of this vector is

$$\|\mathcal{M}_{\mathcal{A}}\{x_i\}\| = \frac{1}{2N}\sqrt{\frac{N-1}{N}} \quad (12)$$

for any $x_i$ in $\mathcal{A}$. Hence, all of the input symbols are mapped to vectors at equal distance from the origin. Moreover, the distance between $\mathcal{M}_{\mathcal{A}}\{x_i\}$ and $\mathcal{M}_{\mathcal{A}}\{x_j\}$ for any $x_i$ and $x_j$ is

$$\|\mathcal{M}_{\mathcal{A}}\{x_i\} - \mathcal{M}_{\mathcal{A}}\{x_j\}\| = \frac{1}{\sqrt{2}N}.$$

Hence, the input symbols are mapped to equidistant vectors. When these two facts are combined it can be deduced that the range space of the alphabet $\mathcal{A}$ is composed of the vertices of a regular polyhedron whose center of mass is at the origin. For instance, for $N = 3$ the the input symbols are mapped to the vertices of an equilateral triangle with center of mass at the origin.

Mapping an observation depends on the a posteriori probabilities of the input symbols given the observation occurred. Hence, we inherently assume that a posteriori probabilities are well known. With this assumption we propose the following definition for $\mathcal{M}_{\mathcal{R}}\{y\}$ for a $y$ in $\mathcal{R}$.

$$\mathcal{M}_{\mathcal{R}}\{y\} \triangleq \sum_{i=1}^{N} \log \frac{(\Pr\{X=x_i|Y=y\})^N}{\prod_{j=1}^{N}\Pr\{X=x_j|Y=y\}}\mathbf{e}_i. \quad (13)$$

If all the input symbols are equally likely then this definition can also be expressed in terms of likelihood functions as follows.

$$\mathcal{M}_{\mathcal{R}}\{y\} = \sum_{i=1}^{N} \log \frac{(\Pr\{Y=y|X=x_i\})^N}{\prod_{j=1}^{N}\Pr\{Y=y|X=x_j\}}\mathbf{e}_i. \quad (14)$$

Notice that this equation describes a sort of log-likelihood ratio. For instance, for $N = 2$ (14) reduces to

$$\mathcal{M}_{\mathcal{R}}\{y\} = \left[\begin{array}{c} \log\frac{\Pr\{Y=y|X=x_1\}}{\Pr\{Y=y|X=x_2\}} \\ \log\frac{\Pr\{Y=y|X=x_2\}}{\Pr\{Y=y|X=x_1\}} \end{array}\right]$$

The definition of the transformation $\mathcal{M}_{\mathcal{R}}\{.\}$ given in (13) restricts the channels on which our representation is applicable. If $\Pr\{X=x_i|Y=y\}$ is zero for an $x_i$ in $\mathcal{A}$ and a $y$ in $\mathcal{R}$ then $\mathcal{M}_{\mathcal{R}}\{y\}$ is not in $\mathbb{R}^N$. Therefore, our representation can not be applicable to those channels if an observation rules an input symbol out, for instance, the binary erasure channel. This is the sole restriction that our representation has.

The range space of the operators $\mathcal{M}_{\mathcal{A}}\{.\}$ and $\mathcal{M}_{\mathcal{R}}\{..\}$ are squeezed into a certain hyperplane of $\mathbb{R}^N$. Let a subset of $\mathbb{R}^N$ be defined as

$$\mathcal{P} \triangleq \{\mathbf{p} \in \mathbb{R}^N : [\begin{array}{cccc} 1 & 1 & \cdots & 1 \end{array}]\mathbf{p} = 0\}$$

Clearly, $\mathcal{P}$ is a $N-1$ dimensional subspace of $\mathbb{R}^N$. More specifically, $\mathcal{P}$ is a hyperplane. For any $x_i$ in $\mathcal{A}$

$$\begin{aligned}[\begin{array}{cccc} 1 & 1 & \cdots & 1 \end{array}]\mathcal{M}_{\mathcal{A}}\{x_i\} &= \frac{1}{2N} - \sum_{j=1}^{N}\frac{1}{2N^2} \\ &= 0.\end{aligned}$$

Therefore, $\mathcal{M}_{\mathcal{A}}\{x_i\}$ is in $\mathcal{P}$ for any $x_i$ in $\mathcal{A}$. Similarly, for any $y$ in $\mathcal{R}$

$$[\begin{array}{cccc} 1 & 1 & \cdots & 1 \end{array}]\mathcal{M}_{\mathcal{R}}\{y\} = $$
$$\sum_{i=1}^{N} \log \frac{(\Pr\{X=x_i|Y=y\})^N}{\prod_{j=1}^{N}\Pr\{X=x_j|Y=y\}} = 0.$$

Hence, $\mathcal{M}_{\mathcal{R}}\{y\}$ is also in $\mathcal{P}$ for any $y$ in $\mathcal{R}$. Therefore, the operators $\mathcal{M}_{\mathcal{A}}\{.\}$ and $\mathcal{M}_{\mathcal{R}}\{.\}$ actually map the detection problem into an $N-1$ dimensional Euclidean space.

After introducing how to map the input symbols and observations to $\mathbb{R}^N$, we have to prove that these transformations satisfy the necessary requirements imposed in the previous section.

**Theorem 1.** *For any $x_i$ in $\mathcal{A}$ and any $y$ in $\mathcal{R}$*

$$\Pr\{X=x_i|Y=y\} = \frac{1}{C}\exp\left(-\|\mathbf{y}-\mathbf{x}_i\|^2\right) \quad (15)$$

*where $\mathbf{y} = \mathcal{M}_{\mathcal{R}}\{y\}$, $\mathbf{x}_i = \mathcal{M}_{\mathcal{A}}\{x_i\}$, and the normalization constant $C$ is*

$$C = \sum_{j=1}^{N}\exp\left(-\|\mathbf{y}-\mathbf{x}_j\|^2\right)$$

*Proof:* Due to the definition of $\mathcal{M}_\mathcal{A}\{x_i\}$ we have

$$\|\mathbf{y} - \mathbf{x}_i\|^2 = \left\|\mathbf{y} - \frac{1}{2N}\mathbf{e}_i + \frac{1}{2N^2}\sum_{j=1}^N \mathbf{e}_j\right\|^2$$
$$= -\frac{(\mathbf{y})_i}{N} + D \quad (16)$$

where $(\mathbf{y})_i$ denotes the $i^{th}$ component of vector $\mathbf{y}$ and $D$ is given by

$$D = \left\|\mathbf{y} + \frac{1}{2N^2}\sum_{j=1}^N \mathbf{e}_j\right\| - \frac{1}{2N^3} + \frac{1}{4N^2}$$

Similarly, $C$ can be expressed as

$$C = \exp(-D)\sum_{j=1}^N \frac{(\mathbf{y})_j}{M}. \quad (17)$$

The right hand side of (15) can be derived using (16) and (17) as

$$\frac{1}{C}\exp\left(-\|\mathbf{y} - \mathbf{x}_i\|^2\right) = \frac{\exp((\mathbf{y})_i/N)}{\sum_{j=1}^N \exp((\mathbf{y})_j/N)} \quad (18)$$

Inserting the definition of $\mathcal{M}_\mathcal{R}\{y\}$ in this equation yields

$$\frac{1}{C}\exp\left(-\|\mathbf{y} - \mathbf{x}_i\|^2\right) = \frac{\frac{\Pr\{X=x_i|Y=y\}}{\left(\prod_{j=1}^N \Pr\{X=x_j|Y=y\}\right)^{1/N}}}{\frac{\sum_{j=1}^N \Pr\{X=x_j|Y=y\}}{\left(\prod_{j=1}^N \Pr\{X=x_j|Y=y\}\right)^{1/N}}}$$
$$= \Pr\{X = x_j|Y = y\}$$

which completes the proof. ∎

We have stated the desired form of the relation between $\Pr\{X|Y\}$ and $\|\mathcal{M}_\mathcal{A}\{X\} - \mathcal{M}_\mathcal{R}\{Y\}\|$ in (1). Moreover, we have explained that the function $f(.)$ should be in a certain form given in (8). The Theorem 1 shows that the operators $\mathcal{M}_\mathcal{A}\{.\}$ and $\mathcal{M}_\mathcal{R}\{.\}$ satisfy these requirements.

The next theorem shows that the operators $\mathcal{M}_\mathcal{A}\{.\}$ and $\mathcal{M}_\mathcal{R}\{.\}$ satisfy (10).

**Theorem 2.** *Let $M$ replicas of an equally likely distributed symbol $X$ transmitted through a memoryless channel and the vector $\mathbf{Y} = [Y_1, Y_2, \ldots, Y_M]$ is received. Then a posteriori probability of $X$ given $\mathbf{Y}$ is*

$$\Pr\{X = x_i|\mathbf{Y} = \mathbf{y}\} = \frac{1}{C}\exp\left(-\left\|\sum_{j=1}^M \mathbf{y}_j - \mathbf{x}_i\right\|^2\right) \quad (19)$$

*where $\mathbf{y}_j = \mathcal{M}_\mathcal{R}\{y_j\}$ [1], $\mathbf{x}_i = \mathcal{M}_\mathcal{A}\{x_i\}$, and the normalization constant $C$ is*

$$C = \sum_{k=1}^N \exp\left(-\left\|\sum_{j=1}^M \mathbf{y}_j - \mathbf{x}_k\right\|^2\right)$$

[1] Notice that $\mathbf{y}$ is a vector in $\mathcal{R}^M$ whereas $\mathbf{y}_j$ is a vector in $\mathbb{R}^N$

*Proof:* Since, the channel is memoryless

$$\Pr\{\mathbf{Y} = \mathbf{y}|X = x_i\} = \prod_{j=1}^M \Pr\{Y_j = y_j|X = x_i\}$$

Using this equation, Bayes' theorem, and the fact that $X$ is identically distributed, the a posteriori probability of $X$ becomes

$$\Pr\{X = x_i|\mathbf{Y} = \mathbf{y}\} = \frac{\prod_{j=1}^M \Pr\{X = x_i|Y_j = y_j\}}{\sum_{k=1}^N \prod_{j=1}^M \Pr\{X = x_k|Y_i = y_i\}}$$
$$= \frac{\exp\left(-\sum_{j=1}^M \|\mathbf{y}_j - \mathbf{x}_i\|^2\right)}{\sum_{k=1}^N \exp\left(-\sum_{j=1}^M \|\mathbf{y}_j - \mathbf{x}_k\|^2\right)} \quad (20)$$

The summation in the numerator can be evaluated as

$$\sum_{j=1}^M \|\mathbf{y}_j - \mathbf{x}_i\|^2 = \sum_{j=1}^M \left(\mathbf{y}_j^T\mathbf{y}_j - \mathbf{y}_j^T\mathbf{x}_i - \mathbf{x}_i^T\mathbf{y}_j + \mathbf{x}_i^T\mathbf{x}_i\right)$$
$$= \left\|\sum_{j=1}^M \mathbf{y}_j - \mathbf{x}_i\right\|^2 - \left\|\sum_{j=1}^M \mathbf{y}_j\right\|^2 + \sum_{j=1}^M \|\mathbf{y}_j\|^2 + (M-1)\|\mathbf{x}_i\|^2$$

The summation in the exponential function in the denominator can be evaluated similarly as

$$\sum_{j=1}^M \|\mathbf{y}_j - \mathbf{x}_k\|^2 = \left\|\sum_{j=1}^M \mathbf{y}_j - \mathbf{x}_k\right\|^2 - \left\|\sum_{j=1}^M \mathbf{y}_j\right\|^2 +$$
$$\sum_{j=1}^M \|\mathbf{y}_j\|^2 + (M-1)\|\mathbf{x}_k\|^2$$

$\|\mathbf{x}_k\|^2$ appearing in this equation can be replaced with $\|\mathbf{x}_i\|^2$ due to (12). Then, inserting these expressions to replace summations inside (20) completes the proof. ∎

### A. Maximum a posteriori decision rule

The maximum a posteriori (MAP) decision rule can be expressed within this geometric framework as follows.

$$\hat{x}_{MAP} = \arg\max_x \Pr\{X = x_i|Y = y\}$$
$$= \arg\max_x \frac{1}{C}\exp(-\|\mathbf{y} - \mathbf{x}_i\|)$$

where $\mathbf{x}_i = \mathcal{M}_\mathcal{A}\{x_i\}$ and $\mathbf{y} = \mathcal{M}_\mathcal{R}\{y\}$. Since, $C$ does not depend on $x$ we can safely get rid of it.

$$\hat{x}_{MAP} = \arg\max_x \exp(-\|\mathbf{y} - \mathbf{x_i}\|)$$
$$= \arg\max_x -\|\mathbf{y} - \mathbf{x}_i\|$$
$$= \arg\min_x \|\mathbf{y} - \mathbf{x}_i\| \quad (21)$$

This equation states that MAP detector should minimize the distance between $\mathcal{M}_\mathcal{A}\{x_i\}$ and $\mathcal{M}_\mathcal{R}\{y\}$, which is an expected result.

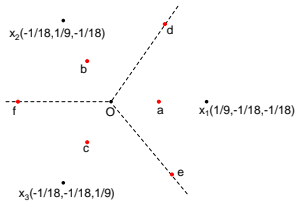

Fig. 1. The plane $\mathcal{P}$ corresponding to the Euclidean geometric representation of the channel described in Section IV-A. The input symbols are mapped to points shown in black whereas the observations are mapped to points shown in red. The dashed lines depict the decision boundaries according to the MAP rule.

## IV. EXAMPLES

In this section we provide some examples of our geometric representation for communication channels. In all of the examples the alphabet consists of three elements. In other words, $\mathcal{A} = \{x_1, x_2, x_3\}$, and $N = 3$. The examples differ in the observation space and/or the probabilistic transition mechanism. As explained in the previous section images of the alphabet and the observation space lie on a hyperplane $\mathcal{P}$. Therefore, we shall only display $\mathcal{P}$ in the illustrations. Since, $N$ is three $\mathcal{P}$ is a two dimensional plane and it is easy to display $\mathcal{P}$. We have to skip the straight forward computations in the examples for the sake of saving space.

### A. A channel with discrete output

The first example is a discrete output channel with input alphabet $\mathcal{A}$ and observation space $\mathcal{R} = \{a, b, c, d, e, f\}$. Let $\Pr\{X|Y\}$ is determined by the following table.

| $\Pr\{X\|Y\}$ | a | b | c | d | e | f |
|---|---|---|---|---|---|---|
| $x_1$ | 0.34 | 0.33 | 0.33 | 0.335 | 0.335 | 0.33 |
| $x_2$ | 0.33 | 0.34 | 0.33 | 0.335 | 0.33 | 0.335 |
| $x_3$ | 0.33 | 0.33 | 0.34 | 0.33 | 0.335 | 0.335 |

The corresponding plane $\mathcal{P}$ is illustrated in Figure 1.

### B. Additive Gaussian noise channel

Suppose that $x_1 = 0$, $x_2 = 1$, and $x_3 = -1$. Let $Y = X + N$ where $N$ is a zero mean normally distributed random variable of variance $\sigma^2$. The observation space $\mathcal{R}$ in this problem is the whole real line. If $X$ is assumed to be equally likely distributed then $\Pr\{X|Y\}$ is given as

$$\Pr\{X = x_i | Y = y\} = \frac{\exp(-\frac{|y-x_i|^2}{2\sigma^2})}{\sum_{j=1}^{3}\exp(-\frac{|y-x_j|^2}{2\sigma^2})}$$

$\mathcal{M}_{\mathcal{R}}\{y\}$ can be computed from (13) after a few steps of derivations as

$$\mathcal{M}_{\mathcal{R}}\{y\} = \frac{1}{2\sigma^2}[\ 2\ \ 6y-1\ \ -6y-1\ ]^T. \quad (22)$$

The corresponding plane $\mathcal{P}$ is shown in Figure 2. Notice that image of $\mathcal{R}$ is a line in this case. The $\mathcal{M}_{\mathcal{R}}\{y\}$ goes upwards in the figure with increasing $y$. The line moves towards right with decreasing $\sigma^2$.

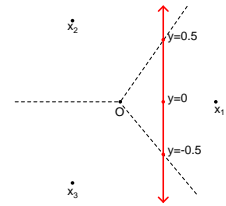

Fig. 2. The plane $\mathcal{P}$ corresponding to the Euclidean geometric representation of the channel described in Section IV-B.

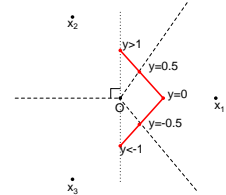

Fig. 3. The plane $\mathcal{P}$ corresponding to the Euclidean geometric representation of the channel described in Section IV-C.

### C. Additive Laplacian noise channel

Consider the very same problem in Section IV-B except $N$ is distributed with a Laplacian distribution with parameter $\lambda$. We have a four piece solution for $\mathcal{M}_{\mathcal{R}}\{y\}$. We omit the derivation and show the resulting $\mathcal{P}$ for a specific $\lambda$ in Figure 3. Notice that all $y > 1$ are mapped to the same point, since, all $y > 1$ results with the same a posteriori probability of $X$.

## V. SUMMARY AND POSSIBLE APPLICATIONS

We have presented an Euclidean geometric representation for the classical detection problem by mapping the hypotheses and observations to $\mathbb{R}^N$. We tried to make this representation useful for the scenarios in which a sequence of symbols is transmitted through the channel. Therefore, our representation may prove useful for applying coding theory to arbitrary channels.

An important possible application might be code design for arbitrary channels. It is well known that increasing the minimum Hamming distance of a code does not always increase the performance of the code. We claim that the distance measure defined below is a good candidate to be optimized while designing codes for arbitrary channels.

$$d_v(\mathbf{c}_1, \mathbf{c}_2) = E\left[\|\mathcal{M}_{\mathcal{R}^M}\{\mathbf{Y}\} - \mathcal{M}_{\mathcal{A}^M}\{\mathbf{c}_2\}\|^2 \,\big|\, \mathbf{X} = \mathbf{c}_1\right]$$

where $\mathbf{c}_1$ and $\mathbf{c}_2$ are two codewords of length $M$ and $E[.|.]$ denotes conditional expectation. Moreover, $d_v(.,.)$ can be computed simply in terms of a symbol-wise distance as

$$d_v(\mathbf{c}_1, \mathbf{c}_2) = \sum_{i=1}^{M} d_s((\mathbf{c}_1)_i, (\mathbf{c}_2)_i)$$

where $d_s(x_i, x_j)$ is $E\left[\|\mathcal{M}_{\mathcal{R}}\{Y\} - \mathcal{M}_{\mathcal{A}}\{x_j\}\|^2 \,\big|\, X = x_i\right]$.

Another possible application might be estimating the performance of non-binary repetition codes which may be serious problem in some channels. Equation (10) may prove useful for this problem.